\documentclass{elsart}

\usepackage{amssymb}
\usepackage{graphicx}
\begin{document}

\begin{frontmatter}



\title{Improving the security of quantum exam against cheating}


\author[BUPT]{Fei Gao\corauthref{cor}},
\corauth[cor]{Corresponding author.} \ead{hzpe@sohu.com}
\author[BUPT]{Qiao-Yan Wen},
\author[CD]{Fu-Chen Zhu}

\address[BUPT]{School of Science, Beijing University of Posts and Telecommunications, Beijing, 100876, China}
\address[CD]{National Laboratory for Modern Communications, P.O.Box 810, Chengdu, 610041, China}

\begin{abstract}
The security of quantum exam [Phys. Lett. A 350 (2006) 174] is
analyzed and it is found that this protocol is secure for any
eavesdropper except for the ``students'' who take part in the
exam. Specifically, any student can steal other examinees'
solutions and then cheat in the exam. Furthermore, a possible
improvement of this protocol is presented.
\end{abstract}

\begin{keyword}
quantum cryptography \sep cryptanalysis \sep entanglement

\PACS 03.67.Hk \sep 03.65.Ud \sep 03.67.Dd
\end{keyword}
\end{frontmatter}

% main text
Cryptography is the approach to assure the secrecy of the data
which is stored or communicated in public environment. From its
beginning the research of cryptography has been progressed along
two directions in parallel. One direction deals with the design of
various schemes to maintain privacy. The other is focused on
analyzing the security of existing protocols, trying to find the
flaws in cryptosystems and improve them. Both directions are
necessary to the development of cryptography. It is also the case
in quantum cryptography \cite{BB84,E91,GRTZ}, where the work of
both scheme designing (e.g. \cite{GGWZ} and references therein)
and security analyzing (e.g. \cite{ZLG01,W04,GGW1,DLZZ}) is
continually proposed.

In a recent paper \cite{N06} a novel protocol called quantum exam
was proposed. In this protocol a teacher Alice wants to organize
an exam with her remotely separated students Bob 1, Bob 2, ... and
Bob $N$. As in a classical exam, all the problems and Bobs'
solutions should not be leaked out and, more importantly, any Bob
cannot obtain other examinees' solutions. However, we find that
the later confidentiality constraint is not perfectly satisfied.
That is, a dishonest Bob can cheating in the exam. In this Letter
we demonstrate this hidden trouble and then present a possible
improvement of the quantum exam protocol.

Let us introduce the quantum exam first. In fact there are two
similar quantum exam protocols presented in Ref.\cite{N06}. We
will take the first one (i.e., the so-called absolutely secure
protocol) as our example. For simplicity we use the same notations
as that in Ref.\cite{N06}. The whole protocol is a little
complicated and here we only describe briefly the related part,
that is, the solution-collecting part (including the
entanglement-sharing process). In this stage Alice generates a
large enough number of ordered nonidentical states
\begin{eqnarray}
|\Phi_p\rangle_{a_p1_p...N_p}=\frac{1}{\sqrt{2}}(|0s_{1_p}s_{2_p}...s_{N_p}\rangle_{a_p1_p...N_p}+|1\overline{s}_{1_p}\overline{s}_{2_p}...\overline{s}_{N_p}\rangle_{a_p1_p...N_p}),
\label{mpe}
\end{eqnarray}
where $s_{n_p}=0$ or 1, $\forall 1\leq n\leq N$, and
$\overline{s}_{n_p}=s_{n_p}\oplus 1$ ($\oplus$ denotes an addition
mod 2). Note that the value of $s_{n_p}$ is known only to Alice.
For each $|\Phi_p\rangle$ Alice stores qubit $a_p$ and sends
qubits $1_p$, $2_p$, ..., $N_p$ to Bob 1, Bob 2, ..., Bob $N$,
respectively. Afterwards, Alice selects a subset of the entangled
states $\{|\Phi_l\rangle\}$ to detect eavesdropping. More
concretely, for each $|\Phi_l\rangle$, Alice measures the qubit
$a_l$ randomly in the basis $B_z$ or $B_x$ and informs every Bob
to perform the same measurement on his corresponding qubit. Then
they check the security of the entanglement distribution process
by verifying
\begin{eqnarray}
j^z_{a_l}=\delta_{0,s_{n_l}}j^z_{n_l}+\delta_{1,s_{n_l}}(j^z_{n_l}\oplus1)
\label{de1}
\end{eqnarray}
for every $n=1,2,...,N$ (when $B_z$ was used) or
\begin{eqnarray}
j^x_{a_l}=\prod^N_{n=1}j^x_{n_l}
\label{de2}
\end{eqnarray}
(when $B_x$ was used), where $j$ represents the measurement
result, $j^z_{a_l}(j^z_{n_l})=\{0,1\}$ corresponding to obtaining
$\{|0\rangle,|1\rangle\}$ and $j^x_{a_l}(j^x_{n_l})=\{+1,-1\}$
corresponding to obtaining $\{|+\rangle,|-\rangle\}$. If there is
no eavesdropping detected, the shared entanglement can be used for
solution-collecting some time later. When needed, Alice and Bobs
measure the remaining ordered $|\Phi_p\rangle$-states
$\{|\Phi_m\rangle_{a_m1_m...N_m}\}$ in basis $B_z$ and record the
outcomes as the secure keys. Let $\{j^z_{a_m}\}$ and
$\{j^z_{n_m}\}$ denote the keys belonging to Alice and Bob $n$,
respectively. Every Bob uses his key as a one-time-pad to encrypt
his solution and sends it to Alice. With the knowledge of
$j^z_{a_m}$ and $s_{n_m}$ Alice can obtain each Bob's key [see
Eq.(\ref{de1})]. Consequently, at the end of the exam Alice will
correctly decrypt Bobs' messages and obtain every Bob's solution.

It can be seen that the solution-collecting process comprises
mainly a multipartite quantum key distribution (MQKD) scheme.
Because the one-time-pad is perfectly secure here, the security of
the whole process lies on that of the key distribution. As we
know, the state $|\Phi_p\rangle_{a_p1_p...N_p}$ has a property of
positive parity, i.e., $j^x_{a_p}\prod^N_{n=1}j^x_{n_p}=+1$. This
wonderful property is subtly employed to detect eavesdropping in
the quantum exam protocol [see Eq.(\ref{de2})]. As a result, the
two constraints Eqs.(\ref{de1}) and (\ref{de2}) can make the exam
secure against various kinds of attacks \cite{N06}. However, we
take notice of another property of
$|\Phi_p\rangle_{a_p1_p...N_p}$, that is, one can entangle an
ancilla $|0\rangle$ into the multipartite entangled state by a
controlled-NOT (CNOT) operation and then disentangle it out from
the obtained state by another CNOT operation. The control qubits
of the two CNOT operations can be any two qubits in
$|\Phi_p\rangle_{a_p1_p...N_p}$ and the target is the ancilla. For
example, for a certain $p$, the multipartite entangled state and
the ancilla compose a composite system
\begin{eqnarray}
|\Gamma\rangle^1=|\Phi\rangle_{a1...N}|0\rangle_g=\frac{1}{\sqrt{2}}(|0s_1s_2...s_N\rangle_{a1...N}|0\rangle_g+|1\overline{s}_1\overline{s}_2...\overline{s}_N\rangle_{a1...N}|0\rangle_g),
\end{eqnarray}
where the subscript $g$ represents the ancilla. If one performs a
CNOT operation $C_{kg}$ (the first subscript $k$ denotes the
control qubit and the second one $g$ denotes the target qubit) on
the qubit $k$ ($1\leq k\leq N$) and the ancilla, the state of the
system changes into
\begin{eqnarray}
|\Gamma\rangle^2=\frac{1}{\sqrt{2}}(|0s_1s_2...s_N\rangle_{a1...N}|s_k\rangle_g+|1\overline{s}_1\overline{s}_2...\overline{s}_N\rangle_{a1...N}|\overline{s}_k\rangle_g).
\end{eqnarray}
Now if one performs another CNOT operation $C_{rg}$ on the qubit
$r$ ($1\leq r\leq N$) and the ancilla, he (she) will obtain
\begin{eqnarray}
|\Gamma\rangle^3&=&\frac{1}{\sqrt{2}}(|0s_1s_2...s_N\rangle_{a1...N}|s_k\oplus
s_r\rangle_g+|1\overline{s}_1\overline{s}_2...\overline{s}_N\rangle_{a1...N}|\overline{s}_k\oplus\overline{s}_r\rangle_g)
\nonumber\\
&=&\frac{1}{\sqrt{2}}(|0s_1s_2...s_N\rangle_{a1...N}|s_k\oplus
s_r\rangle_g+|1\overline{s}_1\overline{s}_2...\overline{s}_N\rangle_{a1...N}|s_k\oplus
s_r\rangle_g) \nonumber\\ &=&|\Phi\rangle_{a1...N}|s_k\oplus
s_r\rangle_g.
\end{eqnarray}
It can be seen that the ancilla is disentangled out from the
multipartite entangled state and, more importantly, the original
state $|\Phi\rangle_{a1...N}$ is left alone. As a result, if an
eavesdropper Eve utilizes the above operations to eavesdrop, she
will introduce no errors. Furthermore, when Eve measures the
ancilla in basis $B_z$ she will obtain $s_k\oplus s_r$ definitely.
Since the value $s_k\oplus s_r$ implies, as described as
following, the correlation of the measurement results of qubits
$k$ and $r$, we call the state $|\Phi\rangle_{a1...N}$
``correlation elicitable''. It can be shown that this property
gives a dishonest Bob the chance to cheat in the exam. Without
loss of generality, suppose the dishonest student is Bob $r$ and
he wants to steal Bob $k$'s solution (maybe Bob $k$ is an
outstanding student), he can adopt the following strategy to
achieve his goal.

(i) For each $p$, Bob $r$ prepares an ancilla $|0\rangle$ and
performs two CNOT operations $C_{k_pg_p}$ and $C_{r_pg_p}$ as
described above when Alice distributes the multipartite entangled
states $\{|\Phi_p\rangle_{a_p1_p...N_p}\}$.

(ii) Bob $r$ measures each ancilla in basis $B_z$ and obtains
$s_{k_p}\oplus s_{r_p}$ with certainty.

(iii) Cooperating with Alice, Bob $r$ executes the legal process
to detect eavesdropping and get key bits. After the actions (i)
and (ii), as analyzed above, all the carrier states
$\{|\Phi_p\rangle_{a_p1_p...N_p}\}$ remain unchanged and no
disturbance is introduced. Therefore, Alice cannot detect the
eavesdropping and Bob $r$ will correctly obtain the intended key
bits $\{j^z_{r_m}\}$.

(iv) Bob $r$ gains Bob $k$'s key bits $\{j^z_{k_m}\}$ by simple
calculation. More specifically, Bob $r$ deletes the data
corresponding to the check states $\{|\Phi_l\rangle\}$ from the
bits $\{s_{k_p}\oplus s_{r_p}\}$, and obtains the remaining
ordered bits $\{s_{k_m}\oplus s_{r_m}\}$, which correspond to the
carrier states $\{|\Phi_m\rangle_{a_m1_m...N_m}\}$ and the key
bits $\{j^z_{r_m}\}$. It should be emphasized that, for a certain
$m$, the measurement outcomes of the ancilla $s_{k_m}\oplus
s_{r_m}$ implies the relation between two key bits $j^z_{k_m}$ and
$j^z_{r_m}$, that is, $j^z_{k_m}\oplus j^z_{r_m}=s_{k_m}\oplus
s_{r_m}$. [From Eq.(\ref{mpe}) we can see that either
$j^z_{k_m}=s_{k_m}, j^z_{r_m}=s_{r_m}$ or
$j^z_{k_m}=\overline{s}_{k_m}, j^z_{r_m}=\overline{s}_{r_m}$
holds.] Therefore, with the knowledge of $\{s_{k_m}\oplus
s_{r_m}\}$ and $\{j^z_{r_m}\}$, Bob $r$ can easily get the key
bits $\{j^z_{k_m}\}$ of Bob $k$ by calculating
$j^z_{k_m}=s_{k_m}\oplus s_{r_m}\oplus j^z_{r_m}$ for each $m$.

(v) Bob $r$ cheats when Alice collects the solutions. Obviously,
with the help of $\{j^z_{k_m}\}$, Bob $r$ can decrypt the message
sent from Bob $k$ to Alice and copy Bob $k$'s solution at will.

By this strategy, a dishonest student can steal any other
examinees' solutions. Moreover, the eavesdropping is not difficult
to realize because it needs only facilities similar to that of the
legal parties. One may argue that, in the above example, if Bob
$r$ is far away from the quantum channel between Alice and Bob $k$
he cannot continually perform the two CNOT operations in a certain
time. In fact there is no need to worry about it. Bob $r$ does not
need to take a round trip between his and Bob $k$'s quantum
channels. He can ask his friend, say Charlie, who stands in Bob
$k$'s channel, to perform the first CNOT operation $C_{k_pg_p}$
and then send the ancilla to him.

There is a fact which should be pointed out. That is, the one who
will legally take part in the protocol is prone to be omitted when
we analyze various attack strategies. In fact, in most MQKD
protocols (e.g. quantum secret sharing, see \cite{YG05} and
references therein), a participant generally has more power to
attack than an outside eavesdropper because the participant can
take advantage of the right to access the carrier state partly and
participate in the process of eavesdropping detection. We call
this kind of attack ``participant attack''. In the quantum exam
protocols, as we can see, the eavesdropping result
$\{s_{k_m}\oplus s_{r_m}\}$ does not seem to have much meaning for
an outside eavesdropper, but it is very useful for a participant
Bob to eavesdrop further. Therefore, as implied in
Refs.\cite{HBB99,KKI99,DLZZ05}, the main goal for the security of
an MQKD should be focused on preventing the dishonest participant
from eavesdropping the information.

Now we discuss how to improve the quantum exam protocol to prevent
this kind of participant attack. To retain the features of the
original quantum exam protocol, our aim is to modify it as little
as possible. Since the fundamental reason of this threat is the
speciality of $|\Phi\rangle_{a1...N}$, i.e., ``correlation
elicitable'', Alice can insert some different check qubits to
detect the above attack. For example, before Alice sends the
sequences to Bobs, she inserts a certain number of single qubits
into each sequence in random positions. All these single qubits
are randomly in one of the states $\{|+\rangle,|-\rangle\}$
\cite{Note}. Note that the positions of the single qubits in these
sequences are different from each other. After all Bobs received
their respective sequences, Alice tells each Bob the positions of
these check qubits and lets him measure them in the basis $B_x$.
Then Alice and Bob check the identity of these qubits. If the
error rate is low enough, they proceed with other steps in the
original protocol to finish the quantum exam. Because, for the
dishonest Bob, both the single qubits and the qubits from
$|\Phi\rangle_{a1...N}$ are in maximally mixed state
$\rho=\frac{1}{2}(|0\rangle\langle0|+|1\rangle\langle1|)$, he
cannot distinguish the check qubits from others. Therefore, when
the dishonest Bob wants to cheat using above strategy, he would
introduce errors with probability $\frac{1}{2}$ once he performs a
CNOT operation on a certain check qubit and his ancilla. As a
result, the improved protocol can stand against the above
participant attack. Furthermore, the main frame of the original
protocol is retained and it follows that the security against
other kinds of attacks (such as measure-resend attack, disturbance
attack, entangle-measure attack, etc. \cite{N06}) still holds.

In conclusion, we show that a dishonest student can cheat in the
quantum exam \cite{N06} and give a possible improvement by
inserting some additional check qubits in each sequence. We
emphasize that the participant attack should not be overlooked
when we discuss the security of a MQKD scheme, which generally
possesses more power in eavesdropping than the attack from
outside.

We thank the anonymous reviewer for helpful comments. This work
was supported by the National Natural Science Foundation of China,
Grant No. 60373059; the Major Research plan of the National
Natural Science Foundation of China, Grant No. 90604023; the
National Laboratory for Modern Communications Science Foundation
of China; the National Research Foundation for the Doctoral
Program of Higher Education of China, Grant No.20040013007; the
Graduate Students Innovation Foundation of BUPT; and the ISN Open
Foundation.


\begin{thebibliography}{00}
\bibitem{Note} Here the role of the states
$\{|+\rangle,|-\rangle\}$ is just to prevent the presented attack.
To aquire more security the original strategy to detect
eavesdropping is still needed. We can also use four states
$\{|0\rangle,|1\rangle,|+\rangle,|-\rangle\}$ to prepare the
inserted single qubits, which can totally ensure the security of
these sequences (similar with that of BB84 protocol \cite{BB84}).
We do not choose the latter choice because we try to retain the
features of the original quantum exam protocol, including its
strategy to detect eavesdropping.
\end{thebibliography}
\end{document}